# Spin-based quantum computing using electrons on liquid helium

S.A. Lyon

*Department of Electrical Engineering, Princeton University, Princeton, NJ 08544*

**Abstract:** Numerous physical systems have been proposed for constructing quantum computers, but formidable obstacles stand in the way of making even modest systems with a few hundred quantum bits (qubits). Several approaches utilize the spin of an electron as the qubit. Here it is suggested that the spin of electrons floating on the surface of liquid helium will make excellent qubits. These electrons can be electrostatically held and manipulated much like electrons in semiconductor heterostructures, but being in a vacuum the spins on helium suffer much less decoherence. In particular, the spin orbit interaction is reduced so that moving the qubits with voltages applied to gates has little effect on their coherence. Remaining sources of decoherence are considered and it is found that coherence times for electron spins on helium can be expected to exceed 100 s. It is shown how to obtain a controlled-NOT operation between two qubits using the magnetic dipole-dipole interaction.

There has been considerable excitement in recent years concerning the possibility of constructing a large quantum computer. Such a device must be capable of manipulating the states of many small quantum systems (typically two-level systems, or qubits) and allowing them to interact in a controlled manner, all the while maintaining a very high degree of coherence.[1,2,3] The difficulty of constructing a quantum computer has led to a variety of ingenious proposals spanning many branches of physics.[4,5,6] Among the various promising approaches are semiconductor-based structures in which electron or nuclear spins are the qubits.[7,8,9,10] However, difficult fabrication and materials problems remain. Here we show that many of the same concepts developed for semiconductor heterostructure architectures can be applied to a different heterostructure system: the spin of electrons floating on the surface of liquid helium (LHe). With the electrons residing in a vacuum several important sources of spin decoherence are eliminated, along with electron trapping, which allows for simpler fabrication and operation.

One necessary architectural attribute of a quantum computer is scalability: that increasing the number of qubits not affect the computer's ability to manipulate them, and that the time scale for their manipulation not increase too rapidly with the number of qubits.[3] Examples of physical systems which show promise for scaling include arrays of trapped ions,[4] optically trapped neutral atoms,[5] Josephson-junctions,[6] and various semiconductor devices.[7-10] Solid state implementations are attractive, since the evolution of semiconductor microelectronics over the last five decades is a compelling demonstration of the power of scaling. As shown below, many of the characteristics of solid state implementations which make them scalable are shared by electron spins on helium.

Recently, several authors have suggested novel architectures using the spatial part of the electron wavefunction bound to liquid He[11,12] or solid Ne[13] as the qubit. However, instead we suggest using the electrons' spin as the qubit, and the devices we envision are closer in spirit to the semiconductor-based quantum computer structures proposed for electron spin qubits.[8-10] We consider a geometry in which the electrons are held at the LHe vacuum interface, rather than in a semiconductor heterojunction, and are controlled by gates a few microns below the He surface.[14,15] The structures are direct adaptations of conventional charge-coupled devices (CCD's) and use the same silicon integrated circuit technology.[16] However, binding the electrons to LHe overcomes some of the technical difficulties associated with semiconductors. In particular: (1) The spin orbit interaction produces an effective magnetic field for electrons moving along a hetero-interface,[17] making it difficult to preserve coherence while directly moving electron spin qubits over large distances in semiconductors. This decoherence mechanism is orders of magnitude weaker for electron spins on LHe, expanding the range of architectures which can be employed. (2) Electron traps and potential fluctuations can make the reliable control of individual electrons at semiconductor interfaces problematical, but they are not a critical issue at the LHe-vacuum interface; (3) The long coherence time of the spins residing in the vacuum above the LHe obviates the need for using the fast but difficult to control exchange interaction for entangling spins. Instead, the slower but readily controlled dipole-dipole interaction can be used. Fabrication of the devices we are suggesting is well within the capabilities of many laboratories. Modern integrated circuits require a level of precision which is at least as challenging as the structures we will consider.

A very schematic diagram of a possible gate arrangement for a quantum computer is shown in Fig. 1. The grey areas represent gates which are arranged like those of a conventional 3-phase CCD.[16] The electrons are indicated by the dots with arrows. The heavy horizontal lines are negatively charged "channel stops." The computer will operate with a single electron confined to each pixel (one horizontal period of the structure between two channel stops). Electrons can be rapidly moved with the gates to different areas of the device using standard CCD clock voltage sequences. With a combination of horizontal and vertical CCD's, it is clear that a highly parallel and scalable architecture is possible.

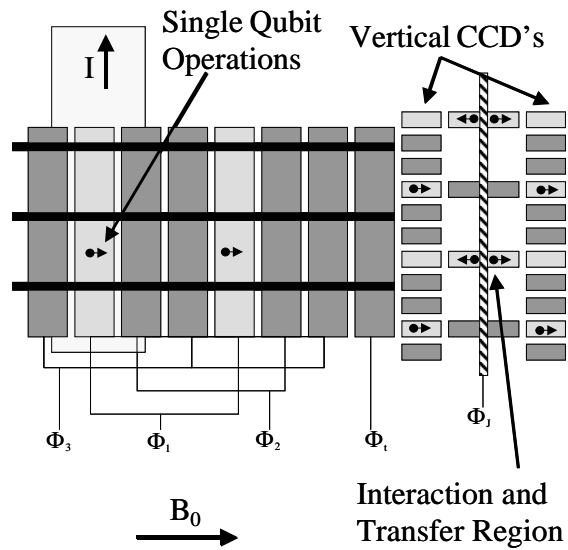

**Figure 1**. A schematic diagram of a possible gate electrode arrangement for a quantum computer. The grey regions are conducting gates held at different potentials ($\Phi_1$ - $\Phi_J$) arranged as a 3-phase CCD. By adjusting the gate voltages the electrons (dots with arrows) can be moved around the device. The light-colored region on the left represents a conductor below the gates through which a current, I, can be run, producing a local lateral magnetic field which combines with the large constant field, $B_0$, to bring the Zeeman transition of the electrons above the wire into resonance with applied microwaves.

Here we will show that all of the operations required for a quantum computer are possible with electrons on the surface of LHe, using reasonable estimates for decoherence rates and other parameters (an applied field, $B_0$, of 0.35 T in the plane of the surface will be assumed, as used in conventional 10GHz electron spin resonance spectrometers). DiVincenzo has compiled a "check list" of five conditions which must be satisfied by a physical implementation of a quantum computer.[18] We will discuss the operation of our devices in the context of this list.

1. There must be a physical embodiment of a qubit.
As discussed above, in the simplest case, the spin of an electron (parallel, $|\uparrow\rangle$, or antiparallel, $|\downarrow\rangle$, to an applied magnetic field) is our qubit. However, there is a considerable advantage to using pairs of electron spins to encode each qubit into a decoherence-free subspace (DFS).[4,19,20,21] In this approach $|0\rangle$ is encoded as $|\uparrow\downarrow\rangle$ and similarly $|1\rangle = |\downarrow\uparrow\rangle$. Since our structures involve spatially varying magnetic fields (both parasitic and intentional), the precise control of a single spin's phase would be difficult. However, with the qubits encoded in a decoherence-free subspace, only the relative phase of the two spins enters. By manipulating the two spins to sequentially follow the same path, the relative phase shift is zero except for temporal variations and imperfections in the paths.

Proposing single electrons, or pairs, as qubits is easy, but isolating and manipulating individual electrons, especially many individual electrons in a large array is very challenging. One of the principal differences between the 2-dimensional (2D) electrons at the LHe-vacuum interface and those in semiconductor heterojunctions is that electrons on LHe maintain high mobilities at very low densities. Whereas the lateral potential profile in semiconductor heterostructures is quite "rough" at low electron densities,[22] leading to practical consequences such as the reduction in the charge-transfer efficiency of Si CCD's for small signals and low temperatures,[23] ultra-high mobility electrons on the surface of bulk LHe have been studied over a density range of $10^5 - 10^9/cm^2$.[24] At these low densities electrons on helium are nondegenerate, even at very low temperatures. They are probably the most perfect example of a *classical*, rather than quantum, 2D electron system.[24] At first this might appear antithetical to a quantum computer, but one of the key difficulties is controlling the quantum entanglement with a high degree of precision. For implementations utilizing electron spins the circuits are substantially simplified if the spins can be held far apart and thus decoupled (both the exchange and dipole-dipole interactions) except when they are being forced to interact.

2. The computations should have a low error rate.
There are two major sources of error. First, the structures used to manipulate the qubits and produce entanglement must do so very precisely. As will be discussed below, we envision using weak interactions so that the time over which fields are acting is long – microseconds to milliseconds. These long times ease the difficulties associated with precision control of spin rotation and entanglement. We have the luxury of using such weak interactions because there will be very little decoherence of the qubits – the second contribution to the error. With the electrons residing in the vacuum above LHe, there are few intrinsic mechanisms which can lead to a loss of spin coherence. To the best of our knowledge there have been no electron spin resonance (ESR) measurements of 2D electrons on the LHe surface; the closest are ESR experiments on electron bubbles in LHe by Reichert and coworkers who obtained a linewidth of $9 \times 10^{-8}$ T (at $B_0 = 0.35$ T), which was still instrumental.[25] Pulsed ESR measurements of the longitudinal spin relaxation time, $T_1$, and coherence time, $T_2$, have not been made.[26]

We have identified several unavoidable sources of spin decoherence which are inherent in the structures described here. One such process is the radiative decay of a spin in its excited state. However, the radiative lifetime of an electron spin in a magnetic field of 0.35 T is of the order of $10^{12}$ s.[27]

A second source of decoherence is the spin-orbit interaction. Silicon and Si/SiGe heterostructures are prime candidates for semiconductor spin-based implementations because of the long intrinsic spin coherence time (>50 ms[28]) for electrons in Si and the availability of isotopically purified Si devoid of nuclear spins. However, confinement of an electron at a heterointerface reduces spin coherence through a combination of the spin-orbit interaction and broken inversion symmetry, the Rashba effect.[17] This effective magnetic field for an electron in a semiconductor can be a significant source of decoherence.[29] As a result even in Si devices the electron-spin coherence time can be reduced to a few microseconds if 2D electrons are free to move in the plane.[30,31] With the electrons in a vacuum above LHe, this spin-orbit term is orders of magnitude smaller. Since the electrons move along the surface of the LHe at velocity $\overline{v}$, and there is a perpendicular electric field, $\overline{E}$, holding them there, they feel an in-plane magnetic field proportional to $\overline{v} \times \overline{E}$. For appropriate parameters ($\overline{v} = 10^5$ cm/s, the electron thermal velocity at 30mK and $\overline{E} = 10^3$ V/cm, the field from the electron's image in the helium) the in-plane field is $B_\parallel \sim 10^{-9}$ T, about six orders of magnitude smaller than at the Si/SiGe interface.[31,32] The direction of the in-plane field depends upon the direction of the velocity so that the spin sees a time-varying magnetic field, and hence it decoheres as the electron scatters, the vacuum equivalent of Dyakonov-Perel relaxation.[29] We are well within the Redfield limit, $\tau \gamma B_\parallel < 1$, where $\tau$ is the momentum relaxation time and $\gamma$ is the electron gyromagnetic ratio. The $T_2$ associated with this process is given by:

$$1/T_2 = \frac{1}{2} \gamma^2 \overline{B_\parallel^2} \tau,$$

where $\overline{B_\parallel^2}$ is the average of the squared in-plane field (the factor of ½ comes from the fraction parallel to $B_0$). On bulk LHe the mobility can be $30 \times 10^6$ cm$^2$/V-s, or a scattering time of about 10 ns,[24] which implies a coherence time of $>10^3$ s. The mobility is lower on thin LHe films, and the $T_2$ from this process can exceed $10^7$ s. Similarly, confined electrons (as in a CCD pixel) scatter more rapidly, and if the confinement is strong enough to freeze the electron into the ground state of its lateral motion this decoherence mechanism is eliminated.

For thin LHe films above a gate another decoherence process arises from Johnson noise currents in the metallic gates producing a fluctuating magnetic field. The currents are larger in low-resistivity metals and become more important as the electron is brought closer to the metallic layer. Thus the most critical region for this decoherence mechanism is where the electrons are held on a thin film of LHe, as in the "interaction region" of Fig. 1 (discussed in more detail below). Sidles and coworkers have analyzed the situation of a spin above a conducting surface,[33] and the coherence time is given by:

$$T_2 = \frac{64\pi}{\gamma^2 \mu_0^2 \sigma} \cdot \frac{d(d+t)}{t} \cdot \frac{1}{kT + \frac{3}{4}\hbar\omega_0 \coth(\frac{\hbar\omega_0}{2kT})},$$

where $\mu_0$ is the magnetic constant, k is Boltzmann's constant, T is the absolute temperature, t is the thickness of the conducting layer with conductivity $\sigma$, $\omega_0$ is the Zeeman frequency in the applied field, and d is the distance of the electron from the surface of the conductor. Assuming a 100Å film of a good metal (resistivity of $1\,\mu\Omega$-cm) and a 400Å thick LHe layer, we estimate the coherence time to be about 1.5 s at 30mK. However, if a degenerately doped semiconductor is used as the metal, the resistivity is higher and the coherence time is $\sim 10^4$ s. The use of such resistive metals will also provide extra damping for the electrons' lateral motion. This will lower their mobility and help keep the electron motion in thermal equilibrium with the bath.

In addition to the currents in a metal, there will be fluctuations in the local spin density. These fluctuations will produce a fluctuating magnetic field acting on the qubits. The thermodynamic fluctuations in the local spin density are given by

$$\langle (n_\uparrow - n_\downarrow)^2 \rangle - (\langle n_\uparrow \rangle - \langle n_\downarrow \rangle)^2$$

where $n_\uparrow$ and $n_\downarrow$ are the local densities of up and down spins, respectively, and the brackets indicate thermodynamic averages. If the fluctuations in up and down spins are assumed to be independent, then the spin density fluctuation reduces to just the fluctuation in the total carrier density, which for a Fermi degenerate system is given by:

$$\langle n^2 \rangle = \left(\frac{3}{\pi^4}\right)^{1/3} \frac{mkT}{\hbar^2} \frac{n^{1/3}}{V},$$

where m is the carrier mass and V is the volume of the local region.[34] If the fluctuations are correlated (for example, if the spin density fluctuates while the local electron density remains constant), then this result is doubled. The characteristic time scale of the fluctuations will be set by the Fermi velocity of a carrier and the distance to the qubit, which with the Redfield formulas[35] gives $T_2 \sim 10^5$ s on a 400Å film of LHe at 30 mK, and longer if the LHe is thicker. Quantum fluctuations may dominate at very low temperatures, but we have not considered them here.

A fifth intrinsic decoherence mechanism we have identified is the coupling of the qubit spins to the nuclear moments of $^3$He atoms. While the atmospheric concentration of $^3$He is low (about 1 part in $10^6$ of $^4$He),[36] at low temperature the $^3$He floats to the surface of the $^4$He.[37] Motional narrowing as the $^3$He atoms diffuse around the surface is somewhat more complicated in this case than in a 3-dimensional system.[38] The electron wave function is peaked some distance above the surface, and we use the results of Neue[39] for the relaxation of dipole-coupled spins confined to two parallel planes. The electron spin coherence time scales linearly with the diffusivity of the $^3$He, which becomes quite large at low temperatures.[40] At 30 mK we calculate a coherence time of over $10^7$ s. $^4$He can be isotopically purified[41] so that the fraction of $^3$He is reduced to $<5\times10^{-13}$, but this appears unnecessary.

It is likely that the actual spin coherence will be controlled by extrinsic factors. An exceptionally high vacuum is attained at mK temperatures, and thus we expect few impurities to accumulate. Paramagnetic defects in the electrode structure and its supports will contribute to decoherence. It is difficult to estimate the quantity and effect of these defects. Such defects will have their spins rapidly relaxed by the metallic layers, reducing their effect on the qubits. If a small number of localized areas are affected, those portions could be replaced by redundant structures in another region of the device. Another extrinsic source of decoherence which is nevertheless difficult to avoid is the ionization produced by cosmic rays. From work with astronomical CCD's it is known that the arrival rate is approximately $10^{-2}/cm^2$-s.[42] However, the ionization rate in LHe is considerably smaller than in semiconductors. Often the true cosmic rays are masked by other environmental sources of radiation. The cosmic ray flux can be reduced by at least two orders of magnitude through very aggressive shielding (a lead room in a deep mine), but such extreme measures should not be necessary.

3. It must be possible to prepare the qubits in a well-defined initial state.
Typically, one would start with a state consisting of all 0's or all 1's. There are several possible ways to accomplish this. The easiest is to simply freeze the spins into their ground state. At 30mK the spins can be frozen into their ground state with an efficiency of $>10^6$:1 for the assumed $B_0$. To accelerate their thermalization the electrons can be stored on a thin LHe layer above a good metal leading to thermalization within a few seconds from the Johnson noise currents.

4. It must be possible to read out the quantum bits.
Reading out the result of a calculation requires qubit measurement, and quantum error correction is most readily implemented via frequent measurements. Probably the simplest way to measure individual electron spins on LHe is to compare the unknown spin with a known one, as has recently been demonstrated in GaAs double quantum dots.[43,44] When two electrons are confined to a small volume (a donor or quantum dot, for example) a singlet-triplet splitting arises from the exchange interaction. If the unknown electron forms a singlet state with the known spin, then the two electrons began with opposite spins. Selectively shifting the phase of one of the electron's spins allows a complete spin measurement. A quantum dot can be constructed for an electron on helium with a locally attractive gate below the surface. Dykman, *et al.* have suggested using submerged metallic posts to form quantum dots.[45] The electron occupancy of a dot can be detected in a variety of ways, for example with a single-electron transistor as demonstrated by Papageorgiou and coworkers.[46]

5. It must be possible to implement single and two-qubit quantum gates.

Single qubit operations can be implemented using conventional pulsed-ESR techniques. Since the spin coherence times are very long, the ESR line is narrow. As noted earlier, the long coherence times allow for long pulses and correspondingly small RF powers. It must be possible to select individual or a subset of qubits to be operated upon, which can be accomplished by setting the RF frequency slightly off resonance as determined by the applied magnetic field. An electric current can be run through wires to generate a local bias field (a fraction of a Gauss) and bring the selected electron spins into resonance. On the left side of Fig. 1 we show, very schematically, a wire (labeled I) running under one of the columns of the CCD structure which would allow all of the spins in that column to be operated upon simultaneously. Spin rotations about the x and y axes can be generated with microwave pulses. Rotations about the z axis can either be generated by a series of rotations about the other two axes, or by locally changing the magnetic field to modify the precession rate of selected spins.

Two-qubit operations, which are at the heart of a quantum computer, are more difficult to design and control. Most semiconductor spin-based quantum computing proposals have employed a variant of the gate-controlled exchange interaction explored by Loss and DiVincenzo.[8] In Fig. 2 we show a structure which will bind two electrons close to one another. The lower part of the figure shows metallic gates with hemispherical bosses (150Å radius) at a separation of 1000Å. The upper part of Fig. 2 shows the calculated one-electron potential energy for an electron held 400Å above the gates by a thin LHe film. The voltages on the gates were all taken to be the same, and adjusted to give zero potential energy for an electron far from the bosses. The electrons are bound to the bosses with a few meV by their image charges in the metal. We have calculated the exchange interaction for the case where the bosses are 30% closer than shown in Fig. 2, and a positive potential is applied to the central gate, $V_J$, to lower the barrier. The exchange coupling can be varied with $V_J$, and it is possible, in principle, to obtain a coupling which allows the "square-root-of-swap" operation to be done in about 1 μs. However, the exchange coupling is extremely sensitive to the parameters of the structure, as has been noted by other authors.[47,48] This is particularly true for electrons on helium since they do not have a small effective mass nor is their repulsion reduced by a large background dielectric constant as for semiconductor quantum dots.

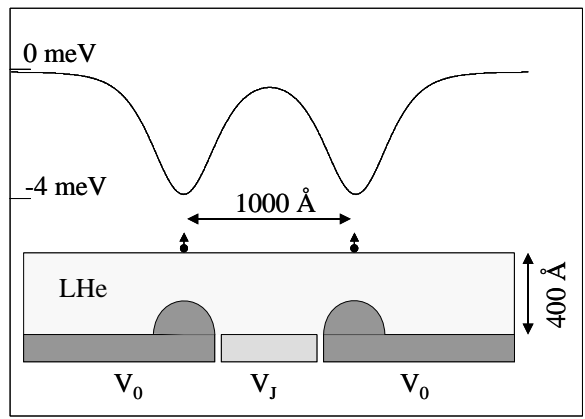

**Figure 2**. A schematic illustration of a structure which can be used to hold 2 electrons (dots with arrows) in close proximity, allowing their spins to interact. The lower part of the figure shows the structure – metallic layers with 2 hemispherical bosses covered by a layer of LHe. The upper part of the figure shows the calculated 1-electron potential energy created by this arrangement of metallic gates with $V_J = V_0$ and chosen to make the electron potential energy approach 0 asymptotically.

Instead of attempting to control the exchange interaction, a more tenable approach is to use the magnetic dipole-dipole coupling between spins. A structure like that shown in Fig. 2 would be used to hold the electrons. For two identical dipoles, $\mathbf{u_1}$ and $\mathbf{u_2}$, the dipole-dipole Hamiltonian is:

$$H_D = \frac{\mathbf{u_1}\cdot\mathbf{u_2}}{r^3} - \frac{3(\mathbf{u_1}\cdot\mathbf{r})(\mathbf{u_2}\cdot\mathbf{r})}{r^5}$$

where $\mathbf{u_1}=g\mu_B S_1$, $\mathbf{u_2}=g\mu_B S_2$, $\mathbf{r}$ is the distance vector from $\mathbf{u_1}$ to $\mathbf{u_2}$, g is the free-electron g-factor (2.00232), $\mu_B$ is the Bohr magneton, and $S_1$ and $S_2$ are the spins of the two electrons.[35] This can be rewritten in the terms of the polar angle, $\theta$, which $\mathbf{r}$ makes with the applied magnetic field, $\mathbf{B_0}$ (defining the z-direction) and the Cartesian components of the spins:

$$H_D = \frac{\mu_0}{4\pi}\frac{g^2\mu_B^2}{r^3}\left\{\frac{(1-3\cos^2\theta)}{2}(3S_{1z}S_{2z}-S_1\cdot S_2) - 3\sin\theta\cos\theta(S_{1z}S_{2z}+S_{1x}S_{2x}) - \frac{3}{2}\sin^2\theta(S_{1x}S_{2x}-S_{1y}S_{2y})\right\}$$

Here the first subscript of S refers to the electron (1 or 2) and the second refers to the direction. The second and third terms, involving $\sin\theta$, are off-diagonal and can be shown to give a correction of the order of the ratio of the dipole field to $B_0$, which is ~$10^{-8}$ in the present situation.[35] Furthermore, with $\theta=0$, these terms vanish. Thus we will only consider the first term in $H_D$.

While the dipole-dipole interaction entangles two spins, by itself it does not make a particularly useful quantum gate. It has been shown that the controlled-NOT (CNOT or quantum exclusive-or) gate together with single-qubit operations is sufficient to make a universal quantum computer.[49] We have found sequences of single-qubit operations, together with the dipole-dipole interaction which produce the CNOT operation. For example, the following sequence produces a phase gate, P:

$$P = \frac{R_{2z}(\pi/2)V_D(\pi/2)R_{1z}(\pi/2)R_{2z}(-\pi/2)V_D(\pi/2)R_{2z}(\pi/2)}{\sqrt{i}} = \begin{pmatrix} 1 & 0 & 0 & 0 \\ 0 & 1 & 0 & 0 \\ 0 & 0 & 1 & 0 \\ 0 & 0 & 0 & -1 \end{pmatrix}.$$

Here $R_{xi}(\phi)$ represents a rotation of spin x by an angle $\phi$ about axis i, $V_D(\Phi)$ represents turning on the dipole-dipole interaction for a time, $\tau$, such that $\Phi = \frac{\mu_0}{4\pi}\frac{g^2\mu_B^2}{r^3}\frac{\tau}{\hbar}$, and the operations are applied right-to-left. The CNOT gate can be obtained from the phase gate with two single-bit rotations[50]

$$\text{CNOT} = R_{1y}(-\pi/2)\,P\,R_{1y}(\pi/2)$$

The time required for an operation will be mainly limited by the magnitude of $H_D$. Single spin rotations with microwave pulses need very little time on this scale. For example, a $\pi$-rotation in a typical pulsed ESR system is about 30ns. However, with 2 electrons in a structure like that shown in Fig. 2, the operation $V_D(\pi)$ will require a time of about 10ms.

This time scales as $1/r^3$, and it may be possible to bring electrons closer together than the 1000Å shown in Fig. 2. The advantage of the weak dipole-dipole interaction is that it can be controlled with high precision, but the price to be paid is slower operation than with exchange coupling. As discussed earlier the long spin coherence times ($T_2 > 100$ s) allow us that luxury. It is important to note that the time for an operation does not increase substantially with the number of qubits since the time to move (or "clock") electrons is in the microsecond range, much shorter than the CNOT time. The high degree of parallelism in the CCD-like structures implies that the overall performance of this system can be very high.

When electrons are bound to thin layers of helium (as in the double-dot structure shown in Fig. 2) their mobility can be significantly lower than on thicker helium. Surface roughness modulates the image potential experienced by the electrons and can lead to trapping.[51] In the CCD channels the liquid helium would be a micron or more thick, and the roughness induced trapping is not significant since the image force at that distance is weak while the CCD fringing fields are large. However, manipulating electrons on a shorter distance scale requires thinner helium, as in the double dot structure, and it is possible that the electrons will become trapped at low temperature. A 1Å thickness variation in a 400 Å helium film will trap an electron with an energy of approximately 0.3K The lateral dimensions of this potential variation will be at least of the order of the helium thickness, implying that the electron can be detrapped with a lateral field of about 5 V/cm. For more severe roughness the thickness of the helium could be varied electrically, as demonstrated in the elegant experiments of Roche and co-workers.[52] Electrons would be transported with a thicker He film, but then the thickness reduced to hold electrons in the devices.

In summary we have shown that a scalable quantum computer using the spin of 2D electrons bound to the surface of LHe shows great promise. The gates needed to move and control the electrons have a similar structure to those used for silicon CCD's, which are currently mass-produced with millions of pixels. Some parts of the structures will require sub-micron patterning, but the fabrication lies well within the bounds of current technology. The magnetic dipole-dipole interaction is suggested for generating entanglement. This gives an operation time of the order of a few ms, which will make precision control of the operations possible. Under other circumstances this period for an operation would be excessive, but according to our estimates of decoherence rates such speeds are tolerable. The construction of a computer with at least tens to hundreds of qubits, scalable to many more, and able to perform tens of thousands of operations before losing coherence appears feasible with current lithographic and materials capabilities.

The author would like to thank G. Csathy, M. Dykman, J.M. Goodkind, M. Shayegan, D.L. Smith, D.D. Smith, D.C. Tsui, and A.M. Tyryshkin for helpful discussions. This work was supported in part by the NSF under grant CCF-0323472, and by the ARO and ARDA under contract W911NF-04-1-0398.

**References:**

1 P. Benioff, Phys. Rev. Lett. **48**, 1581 (1982).

2 D.P. DiVincenzo, Science **270**, 255 (1995).

3 M.A. Nielsen and I.L. Chuang, *Quantum Computation and Quantum Information* (Cambridge University Press, Cambridge, 2000).

4 D. Kielpinski, C. Monroe, and D. J. Wineland, Nat. **417**, 709 (2002).

5 G.K.Brennen, C.M.Caves, P.S.Jessen, and I.H.Deutsch, Phys. Rev. Lett. **82**, 1060 (1999).

6 J.E. Mooij, T.P. Orlando, L. Levitov, L. Tian, C.H. van der Wal, and S. Lloyd, Science **285**, 1036 (1999).

7 B.E. Kane, Nat. **393**, 133 (1998).

8 D. Loss and D.P. DiVincenzo, Phys. Rev. A **57**, 120 (1998).

9 R. Vrijen, E. Yablonovitch, K. Wang, H.W. Jiang, A. Balandin, V. Roychowdhury, T. Mor, and D. DiVincenzo, Phys. Rev. A **62**, 012306 (2000).

10 Mark Friesen, Paul Rugheimer, Donald E. Savage, Max G. Lagally, Daniel W. van der Weide, Robert Joynt, and Mark A. Eriksson, Phys. Rev. B **67**, 121301 (2003).

11 P.M. Platzman and M.I. Dykman, Science **284**, 1967 (1999).

12 A.J. Dahm, J.M. Goodkind, I. Karakurt, and S. Pilla, J. Low Temp. Phys. **126**, 709 (2002).

13 J.M. Goodkind, private communication.

14 Electrons have been bound to other surfaces, including Ne, $^3$He, $H_2$, and LiF, some of which may have advantages, but for definiteness we will consider only $^4$He here.

15 P. Glasson, S. Erfurt Andresen, G. Ensell, V. Dotsenko, W. Bailey, P. Fozooni, A. Kristensen, and M.J. Lea, Physica B **284-288**, 1916 (2000).

16 See for example B.G. Streetmen, *Solid Stare Electronic Devices 4th Edition*, (Prentice Hall, New Jersey, 1995), p. 358.

17 Yu. A. Bychkov and E.I. Rashba, J. Phys. C **17**, 6039, (1984).

18 D. DiVincenzo, cond-mat/9612126.

19 D.A. Lidar, I.L. Chuang, and K.B. Whaley, Phys. Rev. Lett. **81**, 2594 (1998).

20 Lu-Ming Duan and Guang-Can Guo, Phys. Rev. A **57**, 737 (1998).

21 D. Kielpinski, V. Meyer, M.A. Rowe, C.A. Sackett, W.M. Itano, C. Monroe, and D.J. Wineland, Science **291**, 1013 (2001).

22 Z. Wilamowski, N. Sandersfeld, W. Jantsch, D. Többen, and F. Schäffler, Phys. Rev. Lett. **87**, 026401 (2001).

23 A.M Mohsen and M.F. Tompsett, IEEE Trans. Electron Dev. **ED-21**, 701 (1974).

24 C.C. Grimes, Surf. Sci. **73**, 379 (1978), and references therein.

25 J.F. Reichert, and N. Jarosik, R. Herrick and J. Andersen, Phys. Rev. Lett. **42**, 1359 (1979).

26 We are dealing with isolated non-interacting spins, and will use $T_2$ as the coherence time.

27 A. Schweiger and G. Jeschke, *Principles of Pulse Electron Paramagnetic Resonance*, (Oxford University Press, Oxford, 2001).

28 A.M. Tyryshkin, S.A. Lyon, A.V. Astashkin, A.M. Raitsimring,. Phys. Rev. B **68**, 193207 (2003).

29 M.I. Dyakonov and V.I. Perel, Sov. Phys. Solid State **13,** 3023 (1972).

30 Z. Wilamowski and W. Jantsch, Physica E **12**, 439 (2002).

31 A.M. Tyryshkin, S.A. Lyon, W. Jantsch, and F. Schäffler, Phys. Rev. Lett. **94**, 126802, (2005).

32 There will also be a spin-orbit contribution from the small tail of the electrons' wave functions into the He, and their amplitude in He 2P states, but we have not considered that.

33 J.A. Sidles, J.L. Garbini, W.M. Dougherty, S.H. Chao, quant-ph/0004106.

34 L.D. Landau and E.M. Lifshitz, *Statistical Physics $2^{nd}$ edn* (Pergamon, Oxford, 1969).

35 C. P. Slichter, *Principles of Magnetic Resonance $2^{nd}$ edn*, (Springer, Berlin, 1980).

36 W.B. Clark, W.J. Jenkins, and Z. Top, Inter. J. Appl. Rad. Isotopes **27**, 515 (1976).

37 A. F. Andreev, Zh. Eksp. Teor. Fiz. **50**, 1415 (1966)[Sov. Phys. JETP **23**, 939 (1966)].

38 J.-P. Korb, D.C. Torney, and H.M. McConnell, J. Chem. Phys. **78**, 5782 (1983).

39 G. Neue, J. Mag. Res. **78**, 555 (1988).

40 S.K. Lamoreaux, G. Archibald, P.D. Barnes,W.T. Buttler, D.J. Clark, M.D. Cooper, M. Espy, G.L. Greene, R. Golub, M.E. Hayden, C. Lei, L.J. Marek, J.-C. Peng and S. Penttila, Europhys. Lett. **58**, 718 (2002).

41 P.C. Hendry and P.V.E. McClintock, Cryogenics **27**, 131 (1988).

42 A.R. Smith, R.J. McDonald, D.L. Hurley, S.E. Holland, D.E. Groom, W.E. Brown, D.K. Gilmore, R.J. Stover, and M. Wei, SPIE 2002 (available as LBNL-49316).

43  A. C. Johnson, J. R. Petta, J. M. Taylor, A. Yacoby, M. D. Lukin, C. M. Marcus, M. P. Hanson, A. C. Gossard, Nat. **435**, 925 (2005).

44 J. R. Petta, A. C. Johnson, J. M. Taylor, E. A. Laird, A. Yacoby, M. D. Lukin, C. M. Marcus, M. P. Hanson, A. C. Gossard, Science **309**, 2180 (2005).

45 M. I. Dykman, P. M. Platzman, and P. Seddighrad, **67**, 155402 (2003).

46 G. Papageorgiou, P. Glasson, K. Harrabi, V. Antonov, E. Collin, P. Fozooni, P. G. Frayne, M. J. Lea, and D. G. Rees,  Appl. Phys. Lett. **86**, 153106 (2003).

47 A.J. Skinner, M.E. Davenport, and B.E. Kane, Phys. Rev. Lett. **90,** 087901(2003).

48 Belita Koiller, Xuedong Hu, and S. Das Sarma, Phys. Rev. Lett. **88**, 027903 (2002).

49 D.P. DiVincenzo, Phys. Rev. A **51**, 1015 (1995).

50 A. Barenco, C.H. Bennett, R. Cleve, D.P. DiVincenzo, N. Margolus, P. Shor, T. Sleator, J.A. Smolin, and H. Weinfurter, Rhys. Rev. A **52**, 3457 (1995).

51 H.W. Jiang, and A.J. Dahm, Phys. Rev. Lett. **62**, 1396 (1989).

52 P. Roche, G. Deville, K.O. Keshishev, N.J. Appleyard, and F.I.B. Williams, Phys. Rev. Lett. **75**, 3316 (1995).